\documentclass[aps,prl,preprint,superscriptaddress]{revtex4}
\usepackage{graphicx}
\usepackage{dcolumn}
\usepackage{bm}
\usepackage[usenames]{color}
\usepackage{amsmath}
\usepackage{amssymb}
\begin{document}
\title{
Discovery of parity-violating Majorana fermions in a chiral superconductor Sr$_2$RuO$_4$\\
}
\author{Hiroyoshi Nobukane}
\affiliation{Department of Applied Physics, Hokkaido University,
Sapporo 060-8628, Japan}
\author{Akiyuki Tokuno}
\affiliation{Department of Applied Physics, Hokkaido University,
Sapporo 060-8628, Japan}
\author{Toyoki Matsuyama}
\affiliation{Department of Physics, Nara University of Education,
Nara 630-8528, Japan}
\author{Satoshi Tanda}
\affiliation{Department of Applied Physics, Hokkaido University,
Sapporo 060-8628, Japan}
\date{\today}
\begin{abstract}
We found parity-violating Majorana fermions in a chiral superconductor Sr$_2$RuO$_4$. 
The current-voltage curves show an anomalous behavior: The induced voltage is
an \textit{even} function of the bias current.
The magnetic field dependent results suggest the excitation of the Majorana fermions along the closed chiral edge current of the single domain under bias current.  
We also discuss the relationship between a change of the chirality and spontaneous magnetization of the single domain Sr$_2$RuO$_4$.
\end{abstract}
\maketitle
Parity violation provides important insights into physical phenomena of
particle physics and condensed-matter physics. 
As for the parity violation, at least two different mechanisms are known.
The first is due to a quantum anomaly in Dirac fermions of
(2+1)-dimensional quantum electrodynamics~\cite{Semenoff,Matsuyama}. 
Intriguingly, a quantum Hall conductivity in graphene of which
electrical properties are described by Dirac fermions reveals the
parity violation~\cite{Novoselov} and propels
an excellent analogue to (2+1)-dimensional quantum electrodynamics. 
The other mechanism is theoretically predicted to be caused by
(2+1)-dimensional Majorana fermions, where a particle is identified to its own antiparticle. 
However the parity-violating Majorana fermions have never been found yet.
The investigations are very interesting not only in condensed-matter physics,
but also in high-energy physics. Moreover, the application to
topological quantum computation \cite{nayak,dolev,radu} is expected. 

Sr$_2$RuO$_4$~\cite{RMP} is a promising candidate of spin-triplet chiral
$p$-wave superconductor (i.e., electrons are pairing with spin $S=1$ and
orbital angular momentum $L=1$). 
Since the ground states are two-fold degenerate with different
chirality, millimeter-scale Sr$_2$RuO$_4$ is considered to have chiral
domain structures.
Recently, the chiral single domain size has been experimentally
estimated to be $1 \sim 50$ $\mu$m~\cite{kerr, domain, kambara}.
The data of the millimeter-scale Sr$_2$RuO$_4$ should be considered as a
result of ensemble averaging over possible chiral domain configurations.  
Hence transport measurements of samples of the single domain size are very
important in order to study gapless
excitations~\cite{volovik_node,volovik_chiral-edge,stone}, structures and
dynamics of unconventional vortices~\cite{babaev,chung}, 
and a quantum Hall effect~\cite{goryo}.
However transport measurements of the single domain were still limited~\cite{nobukane_SSC}.

In this Letter, we report anomalous current-voltage ($I-V$) characteristics and the magnetic field dependence in a chiral single domain of Sr$_2$RuO$_4$. 
The induced voltage $V$ is an \textit{even} function of the bias current $I$ in the superconducting state. Namely, this violates the parity.
The parity-violating $I-V$ curves show a dependence on the direction of applied magnetic field parallel to $c$ axis.
From the results, we found spontaneous magnetization of about 450 Oe and a change of the chirality of the single domain Sr$_2$RuO$_4$.
Indeed the induced voltage $V$ shows the result of the edge conduction. 
To understand the parity-violating $I-V$ curves with magnetic field dependence, we suggest the low-energy excitation of the right- and the left-handed quasielectrons-holes (Majorana fermions) along the closed chiral edge current of the single domain under bias current.  

The microscale Sr$_2$RuO$_4$ single crystals were prepared in the following procedure.
We synthesized Sr$_2$RuO$_4$ crystals with a solid phase reaction and then determined the crystal structure of Sr$_2$RuO$_4$ and the concentration of impurities.
Next, we selected microscale Sr$_2$RuO$_4$ single crystals from the results of chemical composition and crystallinity. 
These details are described in Ref.~\cite{nobukane_SSC}.
The samples were dispersed in dichloroethane by sonication and deposited on an oxidized Si substrate.  
We confirmed that the dispersed crystals had no boundaries nor
ruthenium inclusions on the sample surface by observing the crystal
orientation using the electron backscatter diffraction pattern \cite{EBSP}. 

\begin{figure}[t]
\begin{center}
\includegraphics[width=0.3\linewidth]{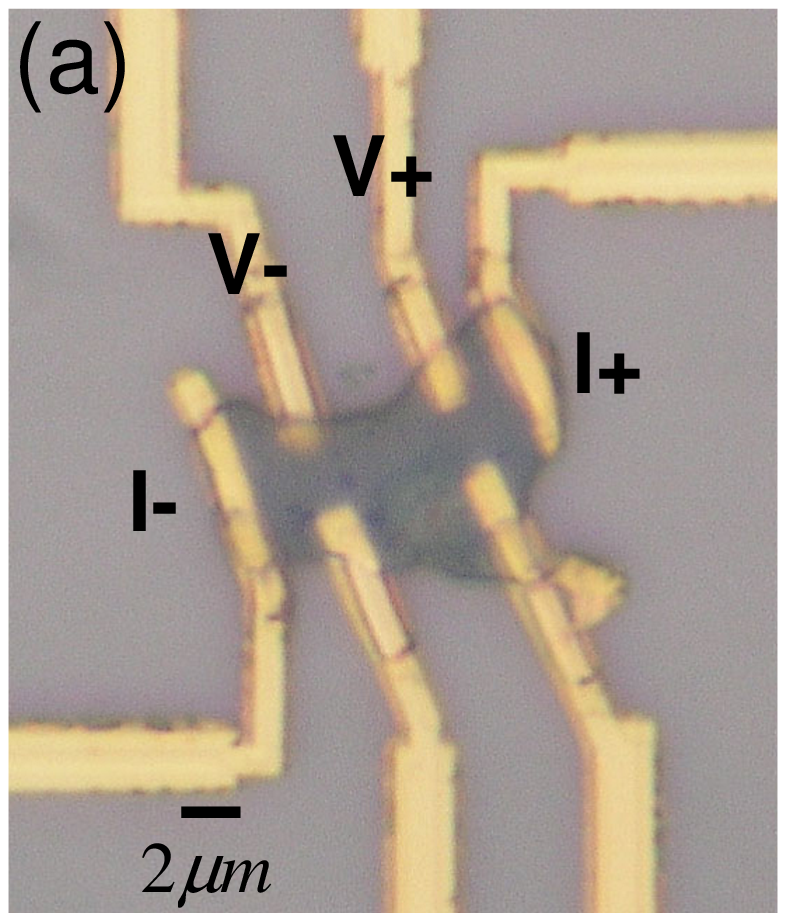}
\includegraphics[width=0.45\linewidth]{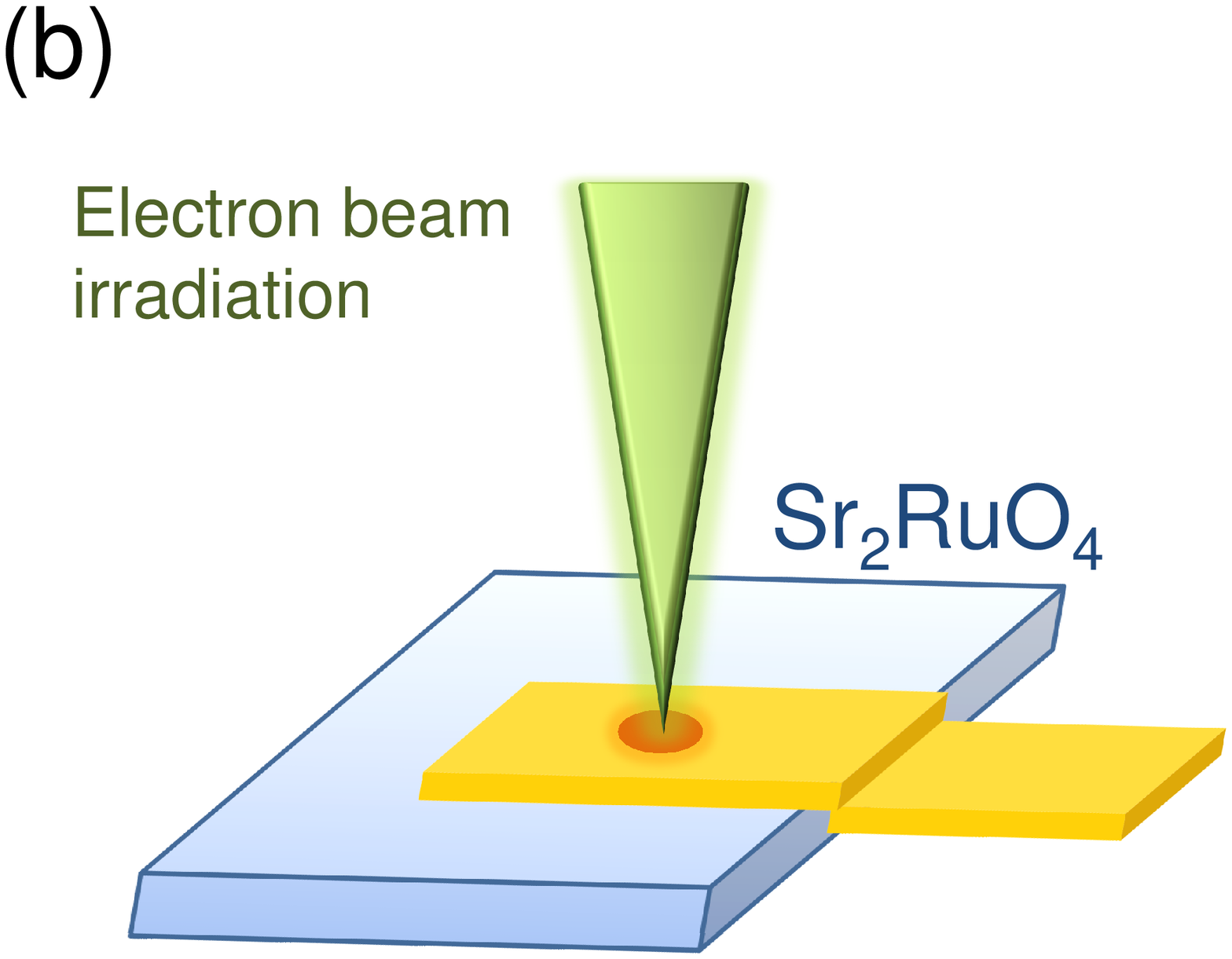}
\includegraphics[width=0.48\linewidth]{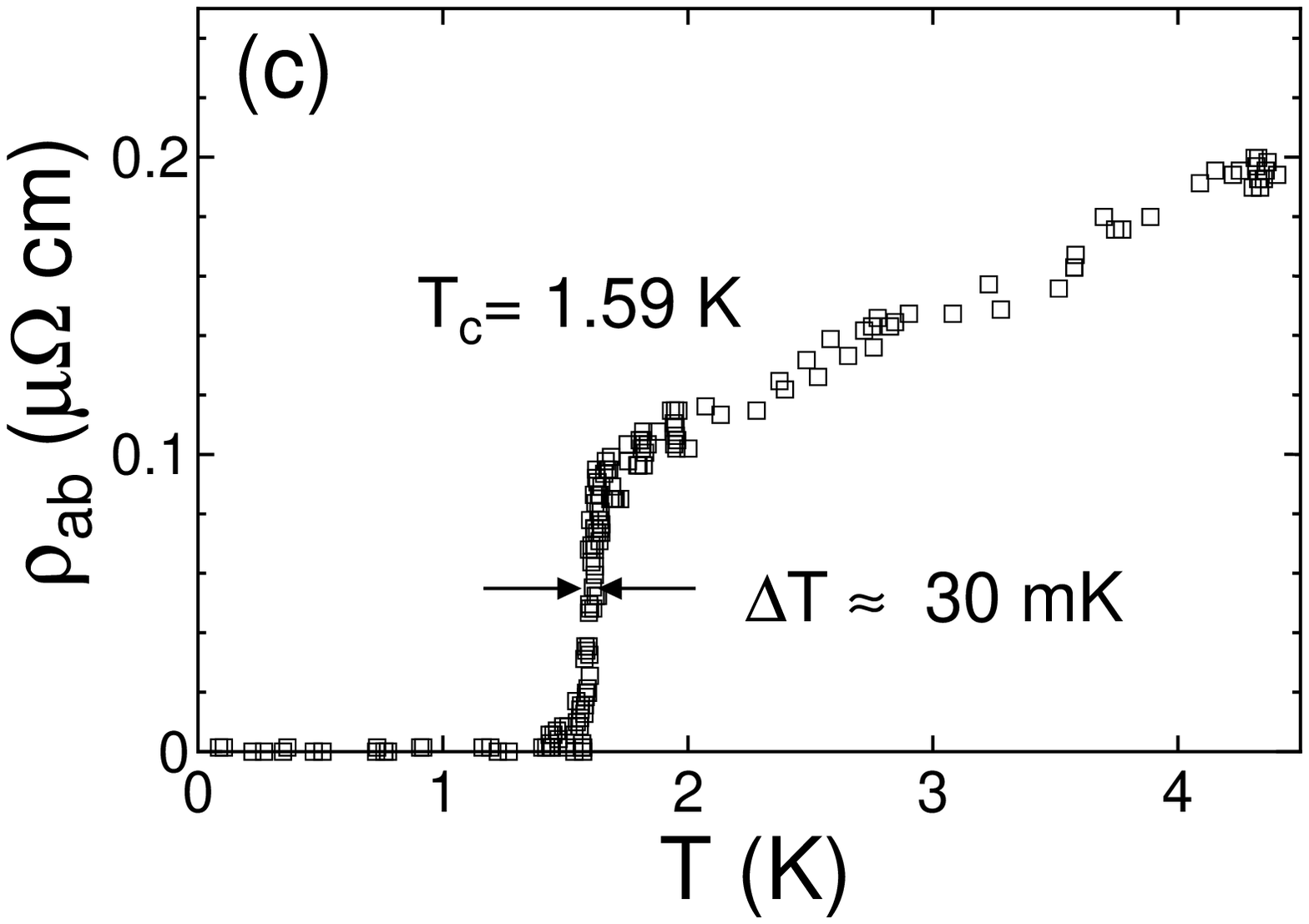}
\includegraphics[width=0.48\linewidth]{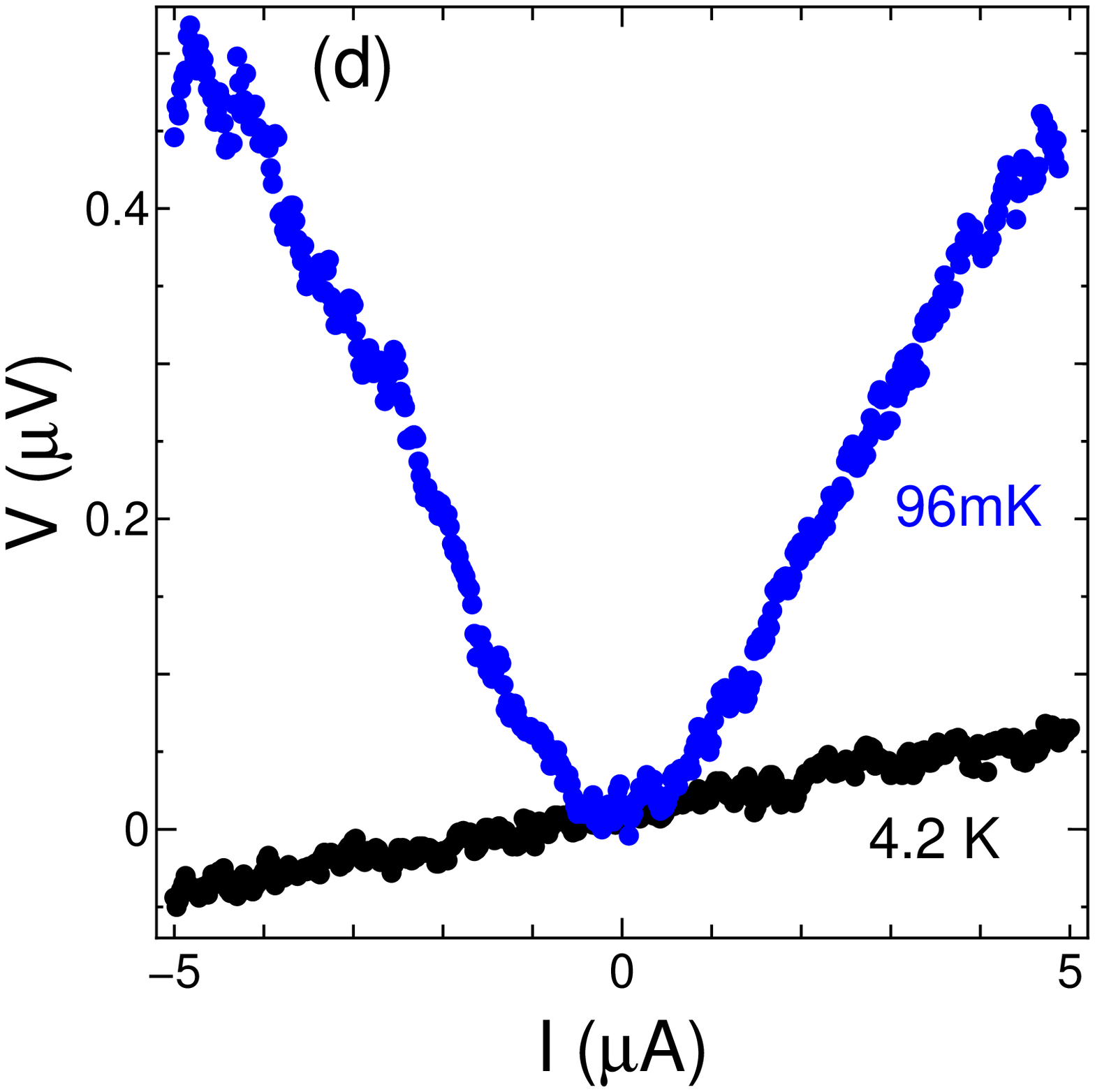}
\caption{
Sample preparation and transport measurements.
(a)A micrograph of a microscale Sr$_2$RuO$_4$ single crystal connected to gold electrodes.
(b)Schematic of the welding using the electron beam irradiation.
(c)Temperature dependence of resistivity of microscale Sr$_2$RuO$_4$ in zero magnetic field. 
(d)Voltage $V$ is plotted as a function of bias current $I$ at $96$ mK and $4.2$ K in zero magnetic field.
The induced voltage shows anomalous behavior which is an \textit{even}
function of the bias current.
Remarkably, the $I-V$ curve at $96$ mK implies the parity violation.
}
\label{figure1}
\end{center}
\end{figure}

On the analysed Sr$_2$RuO$_4$, we fabricated gold electrodes to the sample edges using overlay electron beam lithography.  
Figure~\ref{figure1}(a) shows a micrograph of our samples. 
The sample size is $13.0$ $\mu$m $\times$ $6.67$ $\mu$m $\times$ $0.34$ $\mu$m. The sample electrode spacing is $4.0$ $\mu$m. 
Since the fabricated sample surface may have the insulator surface of
the layer crystals and the residual resist between the sample and the
gold electrodes, it is difficult to form electrical contact. 
Therefore we performed a welding using an electron beam irradiation~\cite{inagaki}. 
Figure \ref{figure1}(b) represents a schematic picture of the electron beam irradiation. 
We heated locally each electrode on the sample for $15$ s with a beam
current irradiation of $2$ $\times$ $10^{-7}$ A.
As the result, we succeeded in greatly reducing the contact resistance
below $1$ $\Omega$ at room temperature.  
Thus the crystal was durably attached with gold electrode to measure
transport properties. 

The measurements were performed in a dilution refrigerator.
All measurement leads were shielded. 
The lead lines were equipped with low pass RC filters (R$=1$ k$\Omega$, C$=22$ nF). 
In the DC measurements, a bias current was supplied by a precise current
source (6220, Keithley) and the voltage was measured with a
nanovoltmeter (182, Keithley) by four-terminal measurements.

Figure~\ref{figure1}(c) shows the superconducting transition temperature
of $T_c=1.59$ K with the transition temperature width of $\Delta T \approx 30$ mK. 
We measured the temperature dependence of the resistivity for bias current $I=\pm$ 1 $\mu$A using the current reversal method.
The resistivity is estimated from the size of $4.0$ $\mu$m (length)
$\times~5.0$ $\mu$m (width) $\times~0.3$ $\mu$m (thickness), and is averaging at temperatures. 
From the results, the temperature dependence of the resistivity in the
microscale Sr$_2$RuO$_4$ was consistent with that of bulk crystals reported in Ref.~\cite{impurity}.
We successfully observed the superconducting properties of the
microscale Sr$_2$RuO$_4$ single crystal.

According to the observed $I-V$ characteristics, we found the parity
symmetry to be violated for the microscale Sr$_2$RuO$_4$ in zero magnetic field below $T_c$.
Figure~\ref{figure1}(d) shows $I-V$ characteristics at $96$ mK and $4.2$ K.
In general, the voltage $V$ in $I-V$ curves is always \textit{odd}
function of bias current $I$, which is a result of parity conservation. 
The $I-V$ curve at $4.2$ K is \textit{odd} function of $I$.
Surprisingly, the $I-V$ curve at $96$ mK shows that $V$ has anomalous components which are \textit{even} function of $I$. 
This implies the parity violation of $I-V$ characteristics.
With decreasing temperature, the anomalous voltage increases in zero
magnetic field.
The resistance $R(=\frac{V_{+}-V_{-}}{I_{+}-I_{-}}$) for $\pm$ 1$\mu$A shows zero at 96 mK, which is consistent with the result of Fig. \ref{figure1}(c).
We note that the measurements do not exceed the critical current density in $ab$ plane  \cite{kambara}.
In order to clarify the nature of the anomalous result of the $I-V$ curve, we have carefully checked possible experimental artifacts, such as measurement errors, thermoelectric effects, and reproducibility of the anomalous effect in several microscale samples, and all of their results supported that the anomaly in the measurement is not caused by extrinsic factors \cite{nobukane_SSC}. 
Thus we can conclude that an intrinsic effect of Sr$_2$RuO$_4$ causes the
parity violation of $I-V$ characteristics. 

\begin{figure}[t]
\begin{center}
\includegraphics[width=0.6\linewidth]{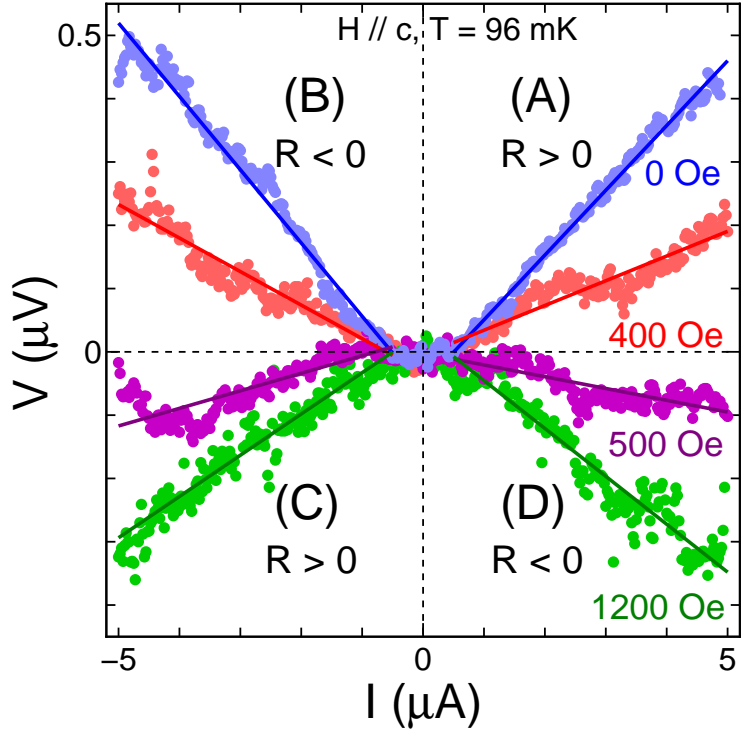}
\caption{
$I-V$ characteristics for several choices of magnetic fields
applied parallel to $c$ axis at $96$mK.
As increasing the magnetic field, the induced voltage changes from
the positive voltage of $V(+I)=V(-I)$ to the negative voltage of
$-V(+I)=-V(-I)$. 
The plotted data are fitted in the region between $I=+(-)0.5$ $\mu$A and
$I=+(-)5.0$ $\mu$A (Solid lines).
We eliminated the offset voltage of about $0.2$ $\mu$V. 
}
\label{figure2}
\end{center}
\end{figure}

Figure~\ref{figure2} shows $I-V$ characteristics for several
magnetic fields applied parallel to $c$ axis at $96$ mK. 
It can be found that all of the $I-V$ curves are symmetric with respect
to the zero bias current: $V(+I)=V(-I)$ or $-V(+I)=-V(-I)$.
Applying the magnetic field from zero magnetic field to $1200$ Oe, the
resistance $R$ on a bias current $I>0$ changes from a positive ($R>0$)
to a negative ($R<0$).
On the other hand, for a bias current $I<0$, a negative $R$ changes to a
positive $R$.
The $I-V$ curve for magnetic field $H=500$ Oe shows zero resistance for
low bias current.
The zero resistance seen in that case 
may be the behavior of the superconductivity.

\begin{figure}[t]
\begin{center}
\includegraphics[width=0.6\linewidth]{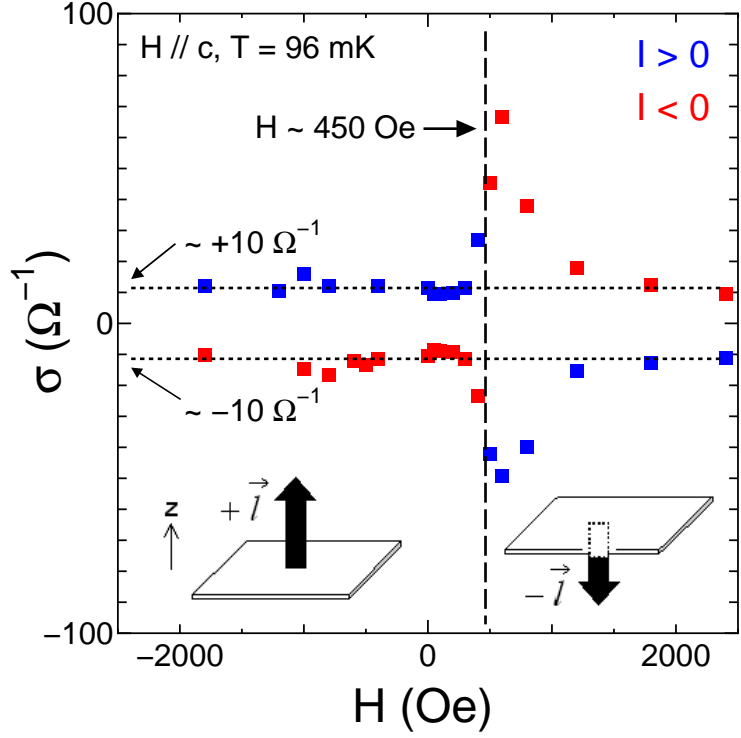}
\caption{
Magnetic field dependence of conductivity for the bias current.
The shift of the infinite conductivity point to $H \sim 450$ Oe and the
exchange of its sign were shown. 
The conductivity of $\pm 10$ $\Omega^{-1}$ is represented by horizontal
dotted lines. 
The insets display the $\pm\vec{l}$ chiral domains in $ab$ plane.
Here we assume that $\pm\vec{l}$ chiral single domain with the broken
time-reversal symmetry is spontaneously formed to the $\pm z$ direction
in zero magnetic field, respectively.
}
\label{figure3}
\end{center}
\end{figure}

To analyze the result of Fig.~\ref{figure2}, we plotted magnetic field
dependence of conductivity for bias current as shown in Fig.~\ref{figure3}. 
The conductivity of conventional superconductors with a magnetic field
has symmetric curves with respect to zero magnetic field and shows
infinite one at zero magnetic field. 
Figure~\ref{figure3}, however, shows that the sample has a finite
conductivity at the zero magnetic field, and a divergent point of
conductivity is shifted to a finite $H\approx 450$ Oe. 
Furthermore, the reflection of the sign of conductivity around the
critical magnetic field $450$ Oe is observed.
In contract, the conductivity is robust against the magnetic field
applied to opposite direction ($H<0$).
In magnetic field away from $450$ Oe, the conductivity asymptotically
reaches $\pm 10$ $\Omega^{-1}$ as shown in horizontal dotted lines of
Fig.~\ref{figure3}. 

To understand the physical origin of the parity-violating $I-V$
characteristics, we firstly discuss the dependence of the chirality of
the single domain of Sr$_2$RuO$_4$ on an external magnetic field. The
chiral domain can explain the shift of the singular point of the
conductivity and the change of its sign in Fig.~\ref{figure3}.  
Now let us suppose that a chiral vector of a single domain, $+\vec{l}$,
spontaneously points to $+z$ direction ($c$ axis) in zero magnetic field
as shown in the left panel of the inset of Fig.~\ref{figure3}. 
The vector $\vec{l}$ represents the pair angular momentum parallel to
$c$ axis.
Assuming that an magnetic field $H_{ex}$ applied to the $-z$ direction
(opposite to $+\vec{l}$) would suppress the chirality of the domain, the
chiral vector of the domain would eventually point to the opposite
direction (i.e., $-\vec{l}$) in a positive effective magnetic field
$H_{eff}>0$, where the effective magnetic field are defined as
$H_{eff}=(H_{ex}-450$ Oe$)$.
In contrast, if the assumption is correct, the magnetic field applied to
the $+z$ direction is expected to affect a change of the chiral domain
with $+\vec{l}$, and the exchange of the sign of the conductivity does
not occur in $H_{eff}<0$.
Actually, according to Fig.~\ref{figure3}, the assumption seems to be
valid.
Our sample size $\sim 10$ $\mu$m is comparable to the single domain
 size $1 \sim 50$ $\mu$m estimated from the experiments~\cite{kerr,domain, kambara}, so that the sample should exhibit a single domain structure.
Thus the microscale Sr$_2$RuO$_4$ itself is the chiral single domain
with spontaneous magnetization and the applied magnetic field causes the
chirality of the domain to change. 

How can we interpret the anomalous $I-V$ curves from the viewpoint
of the peculiar transport phenomena in the chiral single domain? 
In what follows, we focus on an existence of Majorana fermions in
the sample edge of the chiral superconductor Sr$_2$RuO$_4$. 
According to a theoretical prediction, a finite size chiral $p$-wave
superconductor without the time-reversal symmetry has a chiral edge state
which provides gapless excitations \cite{volovik_chiral-edge}, and
the edge excitations of Majorana fermions can be expected \cite{read,zhang}. 
Indeed the supercurrent in the inside of the sample does not contribute to the induced voltage. 
Thus, the transport phenomena at least in a low-energy state may be governed by the edge excitation of the Majorana fermion, and it is possible to explain our results from the viewpoint of the property of Majorana fermions.

\begin{figure}[t]
\begin{center}
\includegraphics[width=1\linewidth]{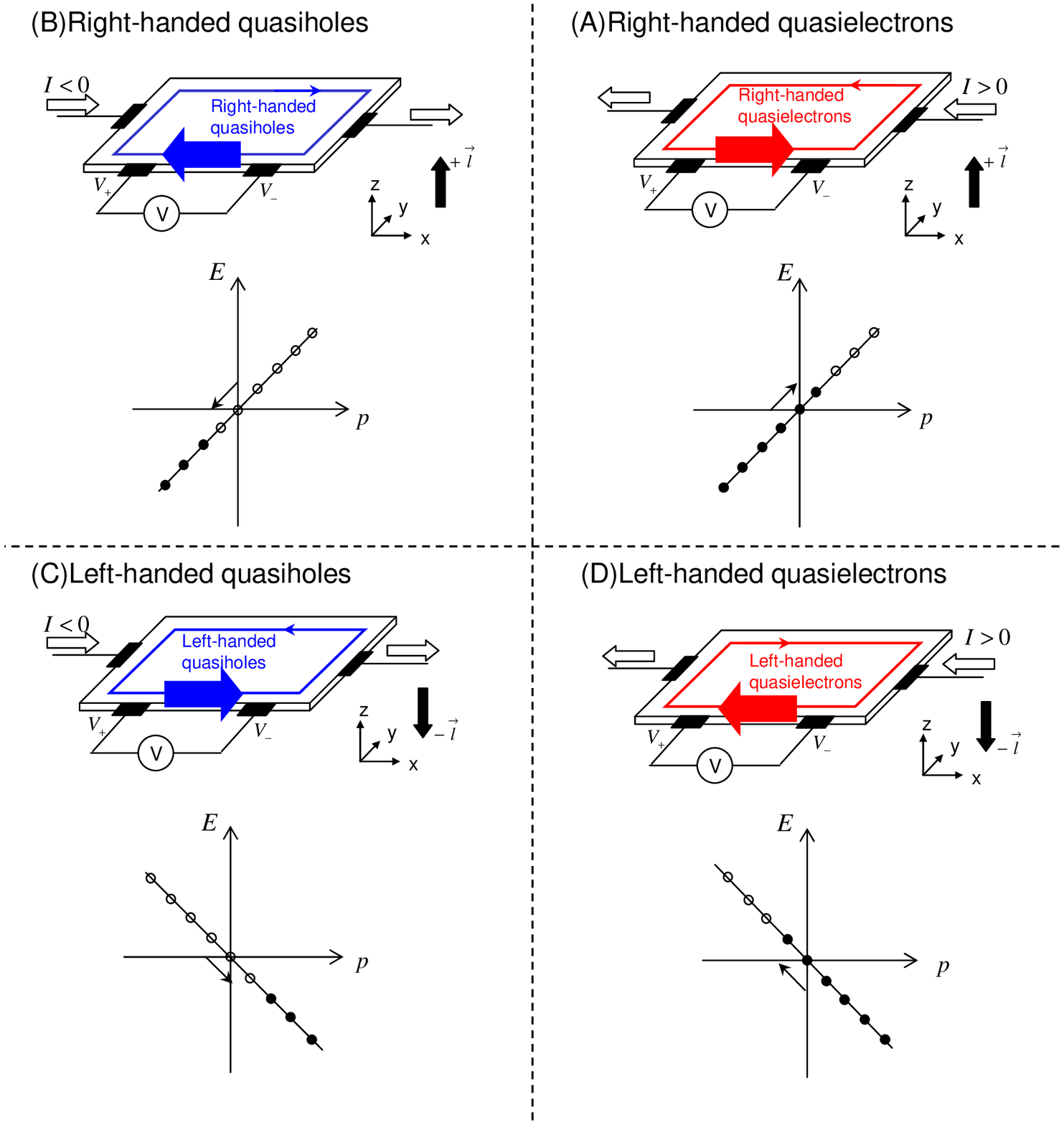}
\caption{
A circulating model of Majorana fermions along the closed chiral edge current of the $+\vec{l}$ or $-\vec{l}$ domain and energy spectrum in chiral edge state.
Under bias current $I>0$ ($I<0$), the right- and the left-handed quasielectrons (quasiholes) are created by the excitation from the Fermi surface (point).
The closed edge of single domain is regarded as a one-dimensional annular line.
The solid and the open circles represent quasielectrons and quasiholes, respectively. 
}
\label{figure4}
\end{center}
\end{figure}

The parity-violating $I-V$ curves can be understood as the result of the selective excitation of chiral quasielectrons-holes (Majorana fermions) along the closed chiral edge current of the single domain under the bias current direction (i.e., $I>0$ or $I<0$). 
As shown in Fig.~\ref{figure4}, we suppose that each region of the $I-V$ curve can be related to the paths of quasielectron or quasihole carrying the current: The panels (A)-(D) respectively correspond to the regions (A)-(D) shown in Fig.~\ref{figure2}.
Here we assume that the chiral edge current of the $+\vec{l}$ domain
flows clockwise for $+z$ direction along the boundary even in the absence of the bias current.
On a bias current $I>0$, when right-handed quasielectrons are created by low-energy excitation from the Fermi surface (point), they circulate along the closed line (Fig. \ref{figure4}(A)). 
On the contrary, for a bias current $I<0$, the right-handed quasiholes created by the excitation circulate oppositely (Fig. \ref{figure4}(B)). 
As a consequence, the anomalous induced voltage of $V(+I)=V(-I)$ was observed. 
Furthermore, similarly to the above explanation, when the $-\vec{l}$ domain, which has the chiral edge current flowing counterclockwise for $+z$, produce the left-handed quasielectrons and quasiholes, the voltage of $-V(+I)=-V(-I)$ is induced as shown in Figs.~\ref{figure4}(C) and (D).
The excitation of quasiholes to the negative energy is considered as the same excitation of quasielectrons to the positive energy in the case of Majorana fermions \cite{zhang}. 
The excitation of the Majorana fermions is determined by the bias current and the direction of the applied magnetic field.
In this manner, related the excitation of the Majorana fermions, the parity-violating $I-V$ curves can be understood. 
It would be an evidence for the parity-violating Majorana fermions in the chiral single domain of Sr$_2$RuO$_4$.

Let us discuss the resistivity in both the normal ($T \geq T_c$) and the superconducting states ($T \ll T_c$).
As shown in Fig. \ref{figure1}(c), the normal resistivity $\rho_{N}$($=R_{N}~\frac{wt}{L}$) is 0.10 $\mu\Omega$ cm at temperature just above $T_c$ from the sample width $w\approx 5$ $\mu$m, the electrode spacing $L\approx 4$ $\mu$m and the thickness $t\approx 0.3$ $\mu$m.
If we assume the width of the state of the edge excitations to be $\lambda_{ab}$, the edge resistivity at 96 mK can be estimated as $\rho_{Edge}= 0.11$ $\mu\Omega$ cm
($=R^\star~\frac{\lambda_{ab}t}{L}$) using the constant resistance $R^\star = 0.1$ $\Omega$ in Fig.~\ref{figure3}, magnetic penetration length $\lambda_{ab}\approx 152$ nm.
The edge resistivity $\rho_{Edge}$ is equivalent to the normal resistivity $\rho_{N}$ at temperature just above $T_c$. 
Namely, the conduction path in the superconducting state becomes the edge.
Thus our result reflects the quasielectrons-quasiholes of the edge current.
From both of the parity-violating $I-V$ curve and the edge conduction,
we can conclude that the anomaly of the induced voltage is the signature of the observation of the Majorana fermions carrying the chiral current along the closed edge of the single domain Sr$_2$RuO$_4$ under bias current.

Finally we discuss physical implications for our results.
As well-known, the Majorana fermions describe the nature of neutrinos
in high-energy physics. 
Thus, the Majorana fermions of a chiral single domain Sr$_2$RuO$_4$ might be regarded as \textit{neutrinos} in condensed-matter systems.
This means a possibility of studying cosmology in the laboratory.
Indeed, among condensed-matter systems, superconductors which is a
charged Bose condensate are most analogous to the cosmology. 
From these reasons, our results will suggest a novel effective action like the Chern-Simons term of ($2+1$)-dimensional quantum electrodynamics, which is hard to find the most relevant action of the theory without a phenomenon.   
In addition to their purely scientific interest, the anomalous transport
phenomena will lead to the potential applications to realize topological
quantum computation of non-Abelian statistics.

In summary, we have observed the parity-violating Majorana fermions in the edge of a chiral single domain of Sr$_2$RuO$_4$ through electric transport.
In the four terminal measurements, the $I-V$ curves show the parity violation with the magnetic field dependence.
The Sr$_2$RuO$_4$ itself provides spontaneous magnetization. The applied magnetic field causes the chirality of the single domain to change. 
We show that the chiral Majorana fermions are excited along the chiral edge current under bias current and detected in the edge of Sr$_2$RuO$_4$.  
\begin{acknowledgements}
The authors thank K. Inagaki, Y. Asano,  K. Ichimura,  K. Yamaya, S. Takayanagi, K. Konno, T. Tsuneta, S. Tsuchiya, J. Ishioka, I. Kawasaki, K. Tenya, H. Amitsuka for experimental help and useful discussions. 
We also are grateful to A. J. Niemi, E. Babaev, M. Chernodub, J. Goryo and N. Hatakenaka. 
This work was supported by a Grant-in-Aid for the 21st Century COE program on ``Topological Science and Technology'' from the MEXT of Japan. 
H.N. acknowledges support from the JSPS. 
\end{acknowledgements}

\newpage
\end{document}